\documentclass[12pt]{article}

\begin{document}

\arraycolsep2pt
\parindent=0pt
\parskip=3pt

\def\BB{\bigskip}
\def\MM{\medskip}

{\Large\bf Wulff shape of equilibrium crystals}

\BB

{\large\bf Salvador Miracle-Sole}

\BB

{\it Centre de Physique Th\'eorique, CNRS, Luminy, Case 907,\hfill\break
F--13288 Marseille, Cedex 9, France}

\BB

{The shape of an equilibrium crystal} is obtained,
according to the Gibbs thermodynamic principle,
by minimizing the total surface free energy
associated to the crystal-medium interface.
To study the solution to this problem, known 
as the {Wulff construction}, is the object of the article.

\BB

{\large\bf Introduction}

\BB

When a fluid is in contact with another fluid, or with a gas,
a portion of the total free energy of the system is
proportional to the area of the surface of contact, and to a
coefficient, the surface tension, which is specific for each
pair of substances. 
Equilibrium will accordingly be obtained when the
free energy of the  surfaces in contact is a minimum. 

When one of the substances involved is anisotropic,   
such as a crystal, 
the contribution to the total
free energy of each element of area depends on its
orientation. 
The minimum surface free energy for a drop of a 
given volume determines then  
the ideal form of the crystal in equilibrium. 

The principle of minimum free energy appears in the fundamental 
work of J.W. Gibbs, ``On the equilibrium of heterogeneous substances''
(1875-1878) where, in particular, he shows 
the role of the anisotropic surface tension for the determination of
the shape of a crystal in equilibrium and discuss the formation of facets.
He also points out the complexities of the actual crystal growth 
and suggest that only very small crystals can have 
an ideal equilibrium form.

 G. Wulff, who made classical experiments on crystal growth, reported
 his results in the paper ``Zur Frage der Geschwindigheit des Wastums
und der Auflosung der Kristallflachen'' (1901), first publisshed in russian 
in 1895. 
His principal conclusions include the celebrated Wulff's theorem:  

``The minimum surface energy for a given volume of a 
polyhedron will be achieved if the distances of its faces 
from one fixed point are proportional to their 
capillary constants.'' 

The term capillary constants was used for the surface tension. 
Wulff himself supported his principle mainly by its  
consequences and 
his attempt at a general proof was incorrect. 
Complete proofs ware later presented 
by M. von Laue, C. Herring, and others. 
An anthology, with comments, of this early work (since Gibbs) 
is presented in the book by Schneer (1970). 

Since the surface tensions depend upon the geometric
distribution of the particles making up the crystal, 
the Wulff theorem establish a relation
between the forms and the structure of crystals, 
and could be used for this study before the 
discovery of X-ray diffraction. 
Although it is not easy to find equilibrium 
crystals in nature one may think 
that an average of the ensemble of forms of a mineral species
approach equilibrium. 

It is only in recent times that equilibrium crystals 
have been produced in the laboratory. 
Most crystals grow under non-equilibrium conditions, 
as predicted by Gibbs, and is a subsequent relaxation 
of the macroscopic crystal that restores the equilibrium. 
This requires transport of material over long distances 
and the time scales can be very long, even for very small 
crystals. 
One has been able to study, however, metal crystals 
in equilibrium of size 1-10 micron.
Equilibration times of a few days were observed. 
A schematic representation of a cubic equilibrium crystal  
is shown in Figure 1. 

A very interesting phenomenon that can be observed on 
equilibrium crystals is the roughening transition.
This transition is characterized by the disappearance of 
the facets of a given orientation from the equilibrium
crystal, when the temperature attains a certain 
particular value. 
Roughening transitions have been found experimentally,  
first, in negative crystals  (i.e., vapor bubbles included in a
crystal) of organic substances. 
The best observations have been made on helium crystals in 
equilibrium with superfluid helium,  
since then the transport of matter and heat is extremely fast.  
Crystals grow to a size of 1-5 mm and relaxation times vary 
from milliseconds to minutes. 
Roughening transitions for three different types of facets 
have been observed
(see, for instance, Wolf et al. 1983). 

\BB

{\large\bf Thermodynamics of equilibrium crystals}

\MM

The problem is to find, at equilibrium, the shape of a droplet of a phase $c$, 
the crystal, inside a phase $m$, called the medium. 

Let ${\bf n}$ be a unit vector in ${\bf R}^d$ and 
consider the situation in which the phases $c$ and $m$
coexist over a plane perpendicular to ${\bf n}$.  
Let $\tau({\bf n})$ be the surface 

\BB


\begin{center}

\setlength{\unitlength}{1mm}
\begin{picture}(40,60)
\thinlines
\put(0,15.3){\line(4,-3){18.5}}
\put(20,55){\line(4,-3){19.25}}
\put(21.5,1.3){\line(4,3){18.5}}
\put(.75,40.5){\line(4,3){19.25}}
\put(0,15.3){\line(0,1){22.7}}
\put(40,15.2){\line(0,1){22.8}}
\put(20,55){\line(0,1){7}}
\put(40,15.2){\line(4,-3){5}}
\put(0,15.3){\line(-4,-3){5}}
\thicklines
\bezier{200}(.75,40.5)(16.25,29.5)(20,29.5)
\bezier{200}(0,38)(15,26)(16.5,23)
\bezier{200}(16.5,23)(18,20)(18.25,1.5)
\bezier{200}(20,29.5)(23.75,29.5)(38.75,40.5)
\bezier{200}(23.5,23)(25,26)(39.75,38)
\bezier{200}(21.5,1.5)(22,20)(23.5,23)
\bezier{30}(18.5,1.5)(20,.5)(21.5,1.5)
\bezier{30}(0,38)(.2,40)(.75,40.5)
\bezier{30}(39,40.5)(39.5,40)(39.75,38)
\put(19,37){$3$}\put(9,17){$1$}\put(29,17){$2$}
\end{picture}

\end{center}

\begin{quote}

{\footnotesize 
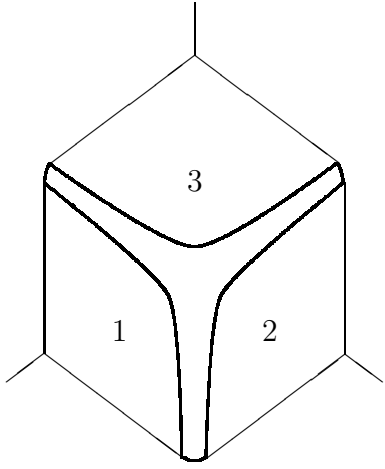
Figure 1. Octant of a cubic equilibrium crystal 
shown in a projection along the (1,1,1) direction.
The three regions 1, 2 and 3 indicate the facets  
and the remaining area represents a curved part 
of the crystal surface.}

\end{quote}


tension,
or free energy per unit area, of such an interface. 
If $B$ is the set of ${\bf R}^d$ occupied
by the phase $c$, and $\partial B$ the boundary of $B$, 
the total surface free energy of the crystal is 
given by
$$
\tau(\partial B)=\int_{\xi\in\partial B}\tau({\bf n}(\xi))
ds_\xi
$$ 
Here ${\bf n}(\xi)$ is the exterior unit normal to
$\partial B$ at $\xi$ and $ds_\xi$ is the element of area at
this point. 
One has to minimize this expression under the 
constraint that the total (Lebesgue) volume $\vert B\vert$ 
occupied by the phase $c$ is fixed. 
Given a set ${\cal W}$, we say that the crystal $B$ has 
shape ${\cal W}$ if after a translation and a dilation it
equals ${\cal W}$.

The solution of this variational problem, known under
the name of Wulff construction, is given by
$$
{\cal W}=\{{\bf x}\in{\bf R}^d : 
{\bf x}\cdot{\bf n}\le\tau({\bf n})\hbox{ for every } 
{\bf n}\in{\bf S}^{d-1}\}
$$ 

{\bf Theorem 1 } {\it 
Let ${\cal W}$ be the just defined Wulff shape for the surface 
tension function $\tau({\bf n})$.
Let $B\subset{\bf R}^d$ be any other region with 
sufficiently smooth 
boundary and the same (Lebesgue) volume as ${\cal W}$.
Then 
$$
\tau(\partial B)\ge\tau(\partial{\cal W})
$$ 
with equality if and only if $B$ and ${\cal W}$ have the 
same shape.}

The proof presented here will be based, 
following Taylor (1987), on geometrical inequalities. 
First, one notices that 
being defined as the intersection of closed 
half-spaces,
the Wulff shape ${\cal W}$ is a closed, bounded convex set, 
i.e., a convex body.  
Among the functions $\tau({\bf n})$, which through the above formula 
define the same shape ${\cal W}$,
there is a unique function having the property
that all planes 
$\{{\bf x}\in{\bf R}^d:{\bf x}\cdot{\bf n}=
\tau({\bf n})\}$, 
associated to all different unit vectors ${\bf n}$,
are tangent to the convex set ${\cal W}$. 
This function is given by
$$
\tau_{\cal W}({\bf n})=
\sup_{{\bf x}\in{\cal W}}({\bf x}\cdot{\bf n})
$$ 
and is called the support function of the convex body 
${\cal W}$.
For an arbitrary function $\tau({\bf n})$
defining the same Wulff shape  ${\cal W}$,
it can be that some of these planes do not
touch the set ${\cal W}$. 
Thus, the function $\tau_{\cal W}({\bf n})$ is 
the smallest function on the unit sphere
which gives the Wulff shape ${\cal W}$. 

Given the surface tension, consider its extension by positive homogeneity,  
$f({\bf x})=\vert{\bf x}\vert\, \tau({\bf x}/ \vert{\bf x}\vert)$.
It turns out that if $ f({\bf x})$ is a convex function on ${\bf R}^d$, then $\tau ({\bf n})$
is the support function of the convex body ${\cal W}$.
This condition is also equivalent to the fact that
the surface tension $\tau$ satisfies a thermodynamic 
stability condition called
the pyramidal inequality, see Messager et al. (1992).

The following Theorem is an extension of the isoperimetric property.  

{\bf Theorem 2 } {\it 
Let ${\cal W}\subset{\bf R}^d$ be a convex body and 
$\tau_{\cal W}({\bf n})$ the corresponding support 
function. 
For any set $B\subset{\bf R}^d$, with sufficiently 
smooth boundary, we have 
$$
\tau_{\cal W}(\partial B)\ge d\ 
\vert{\cal W}\vert^{1/d}\vert B\vert^{(d-1)/d}
$$ 
where $\vert{\cal W}\vert$, $\vert B\vert$, denote the
(Lebesgue) volumes of ${\cal W}$, $B$, respectively,
and $\tau_{\cal W}(\partial B)$ is the  
surface free energy of $\partial B$. 
The equality occurs only when $B$ and ${\cal W}$
have the same shape.}

If $\cal D$ is the closed circle 
of unit radius with center at the origin, then the
corresponding support function $\tau_{\cal D}({\bf n})$ 
is equal to the constant 1, and  the Theorem  reduces to the
isoperimetric property:  
The area $F$ and the length $L$ of any plane domain 
satisfy the inequality
$L^2\ge 4\pi F$.

Theorem 2 is a consequence of the following
Brunn-Minkowski inequality:

{\bf Theorem 3 } 
{\it For non empty compact sets $A,B\subset{\bf R}^d$,
$$
\vert A+B\vert^{1/d}\ge
\vert A\vert^{1/d}+\vert B\vert^{1/d}
$$ 
Moreover, the equality sign holds only when 
$A$ and $B$ are two convex bodies with the same shape
(or one of the sets consists of a single point). }

Given two non-empty sets $A,B\subset{\bf R}^d$
their vector Minkowski sum is defined by
$A+B=\{a+b : a\in A,b\in B\}$. 
A proof of Theorem 3 can be found in the
book by Burago and Zalgaller (1988).  
First one proves by direct computation that the inequality 
holds in the case in which $A$ and $B$ are parallelepipeds 
with sides parallel to the coordinate axis.
The validity of the inequality is then extended by 
induction to all finite unions of such parallelepipeds and,  
finally, to all compact sets by an appropriate limit 
process.

{\bf Lemma }  
{\it Let ${\cal W}$ be a convex body in ${\bf R}^d$. 
Given any set $B\subset{\bf R}^d$, with sufficiently 
smooth boundary, we can express
the functional $\tau_{\cal W}(\partial B)$ as
$$
\tau_{\cal W}(\partial B)
=\lim_{\lambda\to0}{{\vert B+\lambda{\cal W}\vert
-\vert B\vert}\over\lambda}
$$ 
where $\lambda{\cal W}$ denotes the homothetic set 
$\{\lambda{\bf x}:{\bf x}\in{\cal W}\}$. 
If $B\subset{\bf R}^d$ is bounded, the functional
$\tau_{\cal W}$ is well defined.
In particular, this last formula shows that }
$$
\tau_{\cal W}(\partial{\cal W})=d\,\vert{\cal W}\vert
$$ 

{\it Proof}. To prove this lemma, observe that 
if $H({\bf n})$ denotes the half space below a plane 
orthogonal to ${\bf n}$,
and ``dist'' denotes the distance between two sets 
(two parallel planes), 
then, 
from the definition of $\tau_{\cal W}$, 
$$
\tau_{\cal W}({\bf n})={\rm dist}\big(\partial H({\bf n}),
\partial(H({\bf n})+{\cal W})\big)
$$ 
and one can then write
\begin{eqnarray*}
\tau_{\cal W}(\partial B)
&=&\int_{\xi\in\partial B}{\rm dist}\big(\partial H({\bf n}(\xi)),
\partial(H({\bf n}(\xi))+{\cal W})\big)ds_\xi \\
&=&\int_{\xi\in\partial B}\lim_{\lambda\to0}(1/\lambda)
{\rm dist}\big(\xi,\partial(B+\lambda{\cal W})\big)ds_\xi 
\end{eqnarray*} 
This last expression coincides with the formula for 
$\tau_{\cal W}(\partial B)$ 
given in the Lemma, 
and proves its validity. 

{\it Proof of Theorem 2}.
The inequality in Theorem 2 then follows by applying the 
Brunn-Minkowski inequality to 
$\vert B+\lambda{\cal W}\vert$.
This gives
$$
\vert B+\lambda{\cal W}\vert-\vert B\vert\ge
\big(\vert B\vert^{1/d}+\lambda\vert{\cal W}\vert^{1/d}
\big)^d-\vert B\vert\ge
d\,\lambda\,\vert B\vert^{(d-1)/d}\vert{\cal W}\vert^{1/d}
$$ 
which, taking the Lemma into account, ends the
proof of Theorem 2. 

{\it Proof of Theorem 1.}
Let ${\cal W}$ be the Wulff shape corresponding to 
the function $\tau$.
Then
$$
\tau(\partial B)\ge\tau_{\cal W}(\partial B) 
\ge d\ \vert{\cal W}\vert^{1/d}\vert B\vert^{(d-1)/d} 
$$ 
taking the remark into account, together with 
the isoperimetric inequality (Theorem 2).
But, when $B={\cal W}$, we have
$$
\tau(\partial{\cal W})=
\tau_{\cal W}(\partial{\cal W})=d\ \vert{\cal W}\vert
$$ 
Here, the first equality follows from the fact that 
$\tau_{\cal W}({\bf n})\ne\tau({\bf n})$
only for the unit vectors ${\bf n}$ for which 
the planes ${\bf x}\cdot{\bf n}=\tau({\bf n})$
are not tangent to the convex set ${\cal W}$.
The second equality follows from the remark.
Therefore
$$
\tau(\partial  B)\ge \tau(\partial{\cal W})\,
(\vert B\vert/\vert{\cal W}\vert)^{(d-1)/d}
$$ 
which, when $\vert B\vert=\vert{\cal W}\vert$,
gives the inequality stated in Theorem 1. 
The equality in Theorem 1 corresponds to the equality in Theorem 2.
This ends the proof of Theorem 1.

The appearance of a plane facet 
in the equilibrium crystal shape is related 
to the existence 
of a discontinuity in the derivative of the surface
tension with respect to the orientation. 

More precisely, 
let the surface tension $\tau({\bf n})=\tau(\theta,\phi)$, 
for $d=3$, 
be expressed in terms of the spherical coordinates
of ${\bf n}$, 
the vector ${\bf n}_0$ being taken as the $x_3$ axis, 
and assume that it satisfies the stability condition mentioned above.  
Then, $\tau({\bf n})$ is the support function of the Wulff shape, and 
as a natural consequence of this fact
(see Miracle-Sole 1995), it follows: 

{\bf Theorem 4 }
{\it A facet orthogonal to ${\bf n}_0$ appears
in the Wulff shape 
if, and only if, the derivative
$\partial\tau(\theta,\phi)/\partial\theta$
is discontinuous at the point $\theta=0$,
for all $\phi$. 
The facet ${\cal F}\subset\partial{\cal W}$ 
consists of the points ${\bf x}\in{\bf R}^3$
belonging to the plane $x_3=\tau({\bf n}_0)$ and such that, 
for all $\phi$ between $0$ and $2\pi$, } 
$$
x_1\cos\phi+x_2\sin\phi\le
\partial\tau(\theta,\phi)/\partial\theta\,\vert\,_{\theta=0^+}  
$$ 

{\it Proof. }
In terms of the convex function $f({\bf x})=\tau({\bf x}/|{\bf x}|)$, 
the Wulff shape ${\cal W}$ is the set of all
${\bf x}=(x_1,x_2,x_3)\in{\bf R}^3$
such that
$$
x_1 y_1 + x_2 y_2 + x_3 y_3 \le f({\bf y}) 
$$
for every ${\bf y}\in{\bf R}^3$.
If the coordinate axes are placed
in such a way that ${\bf n}_0= (0,0,1)$, 
the plane $x_3=\tau(0)$  
(where $\tau(0)$ is the value of $\tau$ for $\theta=0$)
is a tangent plane to ${\cal W}$.
The facet ${\cal F}$ is the portion of this plane
contained in ${\cal W}$.
Accordingly, the facet ${\cal F}$
consists of the points
$(x_1,x_2,\tau (0))\in{\bf R}^3$
such that
$$ 
x_1 y_1 + x_2 y_2 \le f(y_1,y_2,y_3) - y_3 \tau(0) 
= f(y_1,y_2,y_3) - y_3 f(0,0,1) 
$$
for all ${\bf y}=(y_1,y_2,y_3)$.
Or, equivalently, such that
$$ 
x_1 y_1 + x_2 y_2 \le g(y_1,y_2) 
= \inf_{y_3}\ \big(f(y_1,y_2,y_3)-y_3 f(0,0,1)\big) 
$$
Restricting the infimum to $y_3 = 1/\lambda \ge 0$,
and using the positive homogeneity and the
convexity of $f$, one obtains
$$ 
g(y_1,y_2) = \lim_{\lambda \to 0,\lambda \ge 0} (1/\lambda)\  
\big(f(\lambda y_1,\lambda y_2,1) - f(0,0,1)\big) 
$$
This implies that $g$ is a positively homogeneous 
convex function on ${\bf R}^2$.
Define
$$ 
\mu(\phi) = g(\cos\phi,\sin\phi) 
$$
Taking $\lambda =\tan\theta$, one gets
\begin{eqnarray*} 
\mu(\phi) 
&=& \lim_{\theta\to0,\theta\ge0} (1/\sin\theta)\  
\big(f(\sin\theta \cos\phi,\sin\theta \sin\phi,\cos\theta)
-\cos\theta f(0,0,1)\big) \\
&=& \lim_{\theta \to 0,\theta \ge 0}(1/\theta)\  
\big(\tau(\theta,\phi) - \tau(0)\big) 
=(\partial/\partial\theta)_{\theta=0^+}\tau(\theta,\phi)
\end{eqnarray*}
Similarly
$$
\mu(\phi+\pi) = g(-\cos\phi,-\sin\phi) 
=-(\partial/\partial\theta)_{\theta=0^-}\tau(\theta,\phi)
$$
Both one-sided 
derivatives of $\tau$ exist and,
from the convexity of $g$, it follows that  
$$
\mu(\phi+\pi)\le\mu(\phi) 
$$
Thus, the hypothesis of the discontinuity of the 
derivative $\partial\tau/\partial\theta$, 
at $\theta=0$, implies the strict inequality in the above equation,  
and shows that the convex set ${\cal F}$
has a non-empty interior.

\BB

{\large\bf Interfaces in statistical mechanics} 

\MM

In a first approximation one can model the interatomic 
forces in a crystal by a lattice gas. 
In a typical two-phase equilibrium state there is, 
in these systems, a 
dense component, which can be identified as the crystal 
phase, and a dilute component, which can be identified as the 
vapor phase. 
The underlying lattice structure implies that the 
crystal phase is anisotropic, while this assumption, 
though unrealistic for the vapor phase, should be 
immaterial for the description of the crystal-vapor 
interface. 
As an illustrative example of such systems, 
the ferromagnetic Ising model will be considered.

The Ising model is defined on the $d$-dimensional 
cubic lattice ${\cal L}={\bf Z}^d$, 
with configuration space $\Omega = \{-1,1\}^{\cal L}$. 
The value $\sigma(i)=\pm1$ is the spin at the site $i\in{\cal L}$.  
The occupation numbers,  
$n(i)=(1/2)(\sigma(i)+1)$ $=0$ or $1$,  
give the lattice gas version of this model.
The energy of a configuration
$\sigma_{\Lambda} = \{\sigma(i),i\in\Lambda\}$, 
in a finite box $\Lambda\subset{\cal L}$,
under the boundary conditions ${\bar\sigma}\in\Omega$, 
is
$$
H_{\Lambda}(\sigma_{\Lambda}\mid{\bar\sigma})
= - J \sum_{\langle i,j\rangle\cap\Lambda\not=\emptyset}
\sigma(i)\sigma(j) 
$$ 
where $J>0$, 
$\langle i,j \rangle$ are pairs of nearest neighbor sites,  
and $\sigma(i) = {\bar\sigma}(i)$ when  
$i\not\in\Lambda$.
The partition function, at the inverse temperature 
$\beta=1/kT$, is given by
$$
Z^{\bar\sigma}(\Lambda)
=\sum_{\sigma_{\Lambda}}\exp \big(-\beta
H_{\Lambda}(\sigma_{\Lambda}\mid{\bar\sigma})\big)
$$ 

This model presents, at low temperatures
$T<T_c$, where $T_c$ is the critical temperature, 
two distinct thermodynamic pure phases.
This means two extremal translation invariant Gibbs states, 
which correspond to the limits, when $\Lambda\to\infty$,
of the finite volume Gibbs measures
$$
Z^{\bar{\sigma}}(\Lambda) ^{-1} \exp
\big( -\beta H_{\Lambda}(\sigma_{\Lambda}\mid\bar{\sigma})\big)
$$  
with boundary conditions ${\bar\sigma}$ respectively 
equal to the ground configurations $(+)$ and $(-)$ 
(respectively, ${\bar\sigma}(i) = 1$ 
and ${\bar\sigma}(i) = -1$, for all $i\in{\cal L}$). 
Moreover, they are the unique extremal translation invariant 
Gibbs states of the system (Bodineau, 2006). 
On the other side, if $T\ge T_c$,  
the Gibbs state is unique.

Each configuration inside $\Lambda$ can be described 
in a geometric way by specifying the set of Peierls contours 
which indicate the boundaries between the regions of 
spin $1$ and the regions of spin $-1$.   
Unit square faces are placed 
midway between the pairs of nearest-neighbor sites $i$ 
and $j$, perpendicularly to these bonds, whenever 
$\sigma(i)\sigma(j)=-1$. 
The connected components of this set of faces are 
the Peierls contours. 
Under the boundary conditions $(+)$ and $(-)$, 
the contours form a set of closed surfaces. 
They describe the defects of the considered configuration 
with respect to the ground configurations,  
and are a basic tool for the investigation
of the model at low temperatures. 

In order to study the interface between the two pure phases
one needs to construct a state describing the coexistence 
of these phases. 
To simplify the exposition it will be assumed that $d=3$.
Let $\Lambda$ be a parallelepiped of sides 
$L_1,L_2,L_3$, parallel to the axes,
and centered at the origin of ${\cal L}$, 
and let ${\bf n}=(n_1,n_2,n_3)$ 
be a unit vector in ${\bf R}^3$, such that $n_3\ne 0$. 
Introduce the mixed boundary conditions $(\pm,{\bf n})$, 
for which 
$$ 
{\bar\sigma}(i) = \cases{1  &if\quad $i\cdot{\bf n}\geq 0$ \cr
-1 &if\quad $i\cdot{\bf n}<0$ \cr}  
$$  
These boundary conditions force the system to produce a
defect going trans\-versally through the box $\Lambda$,
a big Peierls contour that can be interpreted as the  
microscopic interface.
The other defects that appear above and below the 
interface can be described by closed contours
inside the pure phases.

The free energy, per unit area, due to the presence of the 
interface, is the surface tension. 
It is defined by
$$
\tau({\bf n})=
\lim_{L_1,L_2\to\infty}\, \lim_{L_3
\to\infty} \, -{{n_d}\over{\beta L_1L_2}}
\ln\, {Z^{(\pm,{\bf n})}(\Lambda)\over 
Z^{(+)}(\Lambda)} 
$$  
In this expression the volume contributions 
proportional to the free energy of the coexisting phases, 
as well as the boundary effects, cancel, and only 
the contributions to the free energy due to the interface 
are left. 

{\bf Theorem 5 } {\it 
The thermodynamic limit $\tau ({\bf n})$, 
of the interfacial free energy per unit area, exists, 
and is a non negative bounded function of ${\bf n}$. 
Its extension by positive homogeneity,  
$f({\bf x})=\vert{\bf x}\vert\, \tau({\bf x}/ \vert 
{\bf x}\vert)$ 
is a convex function on ${\bf R}^3$.}

A proof of these statements has been given by 
Messager et al. (1992) 
using correlation inequalities and, 
in fact, the Theorem holds for a large class of lattice systems.
Moreover, for the Ising model we know, 
from Bricmont et al. (1980), Lebowitz and Pfister (1981), 
and the convexity condition, 
that $\tau ({\bf n})$ is strictly positive for
$T<T_c$ and that it vanishes if $T\ge T_c$.

Consider now the microscopic interface
orthogonal to the direction ${\bf n}_0 = (0,0,1)$.
At low temperatures $T>0$, 
we expect this interface,
which at $T=0$ coincides with the plane $i_3=-1/2$,
to be modified by deformations.
It can be described  by means of its
defects, 
with respect to the interface at $T=0$.
These defects, called walls,
form the boundaries between the smooth plane 
portions
of the interface.
In this way
the interface structure
may then be interpreted as 
a ``gas of walls'' on 
a two-dimensional lattice.

Dobrushin (1972) proved
the dilute character of this gas at low temperatures,
which means that the interface is essentially
flat (or rigid).
The considered boundary conditions yield indeed a
non translation invariant Gibbs state. 
This is known also to be the case for all $T$ less than $T_c^{d=2}$,  
the critical temperature of the two-dimensional Ising model, 
from correlation inequalities (van Beijeren 1975).  
Notice that $J/kT_c^{d=2}=0.44$ is the exact value, and that  
$J/kT_c^{d=3}\sim0.22$ is the estimation in three dimensions. 
It will be seen, in the next Section, that the 
rigidity of the interface is related to the 
formation of a plane facet in the equilibrium crystal.

The same analysis applied to the two-dimensional model
shows a different behavior at low temperatures.
In this case 
Gallavotti (1972) proved that the microscopic 
interface undergoes large fluctuations and does not survive in the
thermodynamic limit. 
The interface is rough and 
the corresponding Gibbs state is translation invariant. 

Coming back to the three-dimensional Ising model, 
where the interface orthogonal to a lattice axis is 
known to be rigid at low temperatures, 
the following question arises:  
At higher temperatures, 
do the fluctuations of this interface become unbounded, 
in the thermodynamic limit, 
so that the corresponding Gibbs state is translation invariant?     

One says then that the interface is rough. 
It is believed that, effectively, the interface becomes rough  
when the temperature is raised, undergoing a roughening 
transition at a temperature $T_R$ strictly smaller than 
the critical temperature $T_c$.
Indeed, approximate methods 
give some evidence for the existence of such a $T_R$ 
and suggest a value near to $J/kT_R=0.41$.  
 
{\bf Remark } 
It has been possible, over the last years, 
to justify the Wulff construction directly from a microscopic theory. 

The first mathematically rigorous proof of the validity of the Wulff construction, 
in the case of the two-dimensional Ising model 
at low temperatures, is due to Dobrushin et al. (1992). 
See also Pfister (1991), for another version of the proof, 
and Miracle-Sole and Ruiz (1994) for a simpler approach in the case of an interface model.

These results show that, using the canonical ensemble, 
where the total number of particles 
(or the total magnetization in the language of 
spin systems) is fixed, 
if the configurations of the system are properly rescaled  
in the thermodynamic limit,
then 
a (unique) droplet of the dense phase, 
immersed in the dilute phase, is formed.
Its shape coincides with the Wulff shape. 
This fact was later extended  
to all temperatures below the critical temperature. 
 
Recently, such a study has also been carried out in the
case of three or more dimensions by Bodineau (1999), 
Cerf and Pitsztora (2000).

\BB

{\large\bf  Wulff shape in statistical mechanics} 

\MM

Consider the surface tension in the Ising model, 
between the positive and negative phases, 
defined as in Theorem 5. 
In the two-dimensional case, 
this function $\tau({\bf n})$ 
has (as shown by Abraham) an exact expression in terms of 
some Onsager's function. 
It follows (as explained in Miracle-Sole 1999) 
that the Wulff shape ${\cal W}$,  
in the plane $(x_1,x_2)$, is given by 
$$
\cosh\beta x_1+\cosh\beta x_2\le\cosh^2 2\beta J/\sinh2\beta J.  
$$ 
This shape reduces to the empty set for $\beta\le\beta_c$, 
since the critical $\beta_c$ satisfies $\sinh2J\beta_c=1$. 
For $\beta>\beta_c$, it is a strictly convex set with smooth 
boundary. 

In the three-dimensional case, 
only certain interface models can be exactly solved. 
Consider the Ising model at zero temperature, 
with boundary condition $(\pm,{\bf n})$. 
Then the ground configurations have only one defect, 
the microscopic interface $\lambda$, 
imposed by this condition, 
and   
$$
\tau({\bf n})=\lim_{L_1,L_2\to\infty}{{n_3}\over{L_1L_2}}
\,\big(E_\Lambda({\bf n})-\beta^{-1}N_\Lambda({\bf n})\big),
$$ 
where $E_\Lambda=2J|\lambda|$ is the energy (all $\lambda$ 
have the same minimal area) and $N_\Lambda$ the number of 
the ground states. 
Every such $\lambda$ has the property of being cut only once by 
all straight lines orthogonal to the diagonal plane 
$i_1+i_2+i_3=0$,   
provided that $n_k>0$, for $k=1,2,3$.  
Each $\lambda$ can then be described by an integer function defined on 
a triangular plane lattice, 
the projection of the cubic lattice ${\cal L}$ 
on the diagonal plane. 
The model defined by this set of admissible microscopic interfaces 
is precisely the TISOS model, introduced by Nienhuis et al. (1989).   
A simi\-lar definition can be given for the BCSOS model that 
describes the ground configurations on the body-centered cubic lattice
(see van Beijeren 1977 and Kotecky, Miracle-Sole 1987). 

{}From a macroscopic point of view,
the roughness or the rigidity of an interface should be apparent
when considering the shape of the equilibrium crystal 
associated with the system.
A typical equilibrium crystal at low 
temperatures has smooth plane facets 
linked by rounded edges and corners. 
The area of a particular facet decreases as the 
temperature is raised and the facet finally disappears 
at a temperature characteristic of its orientation.
It can be argued  
that the disappearance of the facet corresponds to 
the roughening transition of the interface
whose orientation is the same 
as that of the considered facet.

The exactly solvable interface models mentioned above,
for which the function $\tau({\bf n})$ 
has been exactly computed,
are interesting examples of this behavior, 
and provide a valuable information on several aspects 
of the roughening transition. 
This subject has been reviewed in  
Kotecky (1989) and Miracle-Sole (1999).
For example, Figure 1 shows the shape predicted by the TISOS 
model.

For the three-dimensional Ising model at positive temperatures, 
the description of the microscopic interface, for any 
orientation ${\bf n}$, 
appears as a very difficult problem.   
It has been possible, however, to analyze 
the interfaces which are very near 
to the particular orientations ${\bf n}_0$,  
discussed in the precedent Section.
This analysis can be used to determine
the shape of the facets in a rigorous way. 

The step free energy plays a role 
in the facet formation, as shown in Theorems 6 and 7, below.
It is defined as the free energy 
associated with the introduction of a step of height 1   
on the interface,  
and can be regarded as an order parameter
for the roughening transition. 
Let $\Lambda$ be, as in Section 2, a parallelepiped of sides 
$L_1,L_2,L_3$, parallel to the axes, 
centered at the origin, 
and introduce the 
$({\rm step},{\bf m})$
boundary conditions, associated to the unit vectors
${\bf m} = (\cos\phi,\sin\phi)\in{\bf R}^2$, by
$$ 
{\bar\sigma}(i) = \cases{1  & if\ \ $i_3>0$\ \ or if\ \ $i_3=0$\ \ and\ \ 
$i_1m_1+i_2m_2\ge0$,\cr -1  &otherwise. \cr}  
$$ 
Then, the step free energy, per unit length, for a step
orthogonal to ${\bf m}$ (with $m_2>0$) 
on the horizontal interface, is   
$$ 
\tau^{\rm step}(\phi) =
\lim_{L_1\to\infty}\lim_{L_2\to\infty}\lim_{L_3\to\infty}
- {{\cos\phi}\over{\beta L_1}}\  
\ln\  {{Z^{{\rm step},{\bf m}}(\Lambda)}\over 
{Z^{\pm,{\bf n}_0}(\Lambda)}} 
$$ 

A first result concerning the facet formation in the Wulff shape,   
was obtained by Bricmont et al. (1986), by proving 
a correlation inequality which establish
$\tau^{\rm step}(0)$ 
as a lower bound, strictly positif for $T<T_c^{d=2}$,       
to the one-sided derivative
$\partial\tau(\theta,0)/\partial\theta$ at $\theta=0^+$
(the inequality extends to $\phi\ne0$), 
Thus, when $\tau^{\rm step}>0$, a facet is expected, 
according to Theorem 4. 

Using the perturbation theory of the horizontal interface,  
it is possible to study also the 
microscopic interfaces associated with the 
$(\hbox{\rm step},{\bf m})$ boundary conditions. 
When considering these configurations, 
the step may be viewed as an additional defect
on the rigid interface described in Section 2. 
It is, in fact,
a long wall going from one side to the other side 
of the box $\Lambda$. 
The step structure at low temperatures can then be 
analyzed with the help of a new cluster expansion. 
As a consequence of this analysis we have 
the following theorem.

{\bf Theorem 6 } {\it
If the temperature is low enough,  
i.e., if $\beta J\ge c_0$, where $c_0$ is a suitable constant,  
then the step free energy, $\tau^{\rm step}(\phi)$,
exists, is strictly positive, and 
extends by positive homogeneity to a strictly convex function.
Moreover, $\beta\tau^{\rm step}(\phi)$ is an
analytic function of $\zeta=e^{-2J\beta}$, 
for which an explicit convergent series expansion can be found. }

Using the above results on the step structure, 
similar methods allow us to evaluate
the increment in surface tension of an interface 
titled by a very small angle $\theta$ 
with respect to the rigid horizontal interface. 
This increment can be expressed in terms of the 
step free energy and one obtains the following relation. 
                                           
{\bf Theorem 7 } {\it 
For $\beta J\ge c_0$, we have } 
$$
\partial\tau(\theta,\phi)/\partial\theta\,\vert\,_{\theta=0^+}
= \tau^{\rm step}(\phi).   
$$ 

This relation, together with Theorem 4,
implies that one obtains the shape of the facet 
by means of the two-dimensional Wulff construction
applied to the step free energy.
The reader will find a detailed discussion on these points, 
as well as the proofs of Theorems 6 
and 7, in Miracle-Sole (1995). 

{}From the properties of $\tau^{\rm step}$ stated in Theorem 6 
it follows that the Wulff equilibrium crystal 
presents well defined boundary lines, smooth and  
without straight segments, between
a rounded part of the crystal surface and the plane facets parallel
to the three main lattice planes.

It is expected, but not proved, that at a higher temperature, 
but before reaching the critical temperature, 
the facets associated with the Ising model undergo a 
roughening transition. 
It is then natural to believe that the equality in Theorem 7  
is true for any $T$ lower than $T_R$,   
and that for $T$ higher than $T_R$,  
both sides in this equality vanish,
and thus, the disappearance of the facet is involved.
However, the condition that the temperature 
is low enough is needed in the proofs of Theorems 6
and 7. 

\newpage

{\large\bf References}

\MM

Beijeren, H. van (1975): Interface sharpness in the Ising model. 
Commun. Math.Phys. {\bf 40}, 1-6. 

Beijeren, H. van, (1977):
Exactly solvable model for the roughening transition 
of a crystal surface, 
Phys. Rev. Lett. {\bf 38}, 993-996. 

Bodineau, T. (1999):
The Wulff construction in three and more dimensions,
Commun. math. Phys. {\bf 207}, 197-229. 

Bodineau, T. (2006):
Translation invariant Gibbs states for the Ising model,  
Probab. Theory Related Fields {\bf 135}, 153-168. 

Bricmont, J., Lebowitz, J.L., Pfister, C.E. (1980):
On the surface tension of lattice systems, 
Ann. Acad. Sci. New York {\bf 337}, 214-223. 

Bricmont, J., El Mellouki, A., Fr\"ohlich, J. (1986):
Random surfaces in statistical mechanics: roughening, rounding, wetting,...
J. Stat. Phys. {\bf 42}, 743--798. 

Burago, Yu.D., Zalgaller, V.A. (1988):
Geometric inequalities,  
Grundiehren der math. Wissenschaften, 
vol. 43, Springer, Berlin. 

Cerf, R., Pitsztora, A. (2000): 
On the Wulff crystal in the Ising model, 
Ann. Probab. {\bf 28}, 945-1015.

Dobrushin, R.L. (1972):
Gibbs state describing the coexistence of phases
for a three dimensional Ising model, 
Theory Probab. Appl. {\bf 17}, 582-600.

Dobrushin, R.L., Kotecky, R., Shlosman, S.B. (1992):
The Wulff Construction: a Global Shape from Local Interactions,
American Mathematical Society, Providence. 
Preprint, 
http://www.cpt.univ-mrs.fr/dobrushin/DKS-book.pdf  
 
Gallavotti, G. (1972):
The phase separation line in the two-dimensional Ising model,
Commun. Math. Phys. {\bf 27}, 103-136. 

Kotecky, R. (1989):
Statistical mechanics of interfaces and equilibrium crystal shapes.  
In: IX International Congress of Mathematical Physics, 
Simon, B. et al. (Eds.),  
pp. 148-163, Adam Hilger, Bristol.

Kotecky, R., Miracle-Sole, S. (1987):
Roughening transition for the Ising model on a bcc lattice. 
A case in the theory of ground states, 
J. Stat. Phys. {\bf 47}, 773-799. 

Kotecky, R., Miracle-Sole, S. (1987): 
A roughening transition indicated by the behaviour of ground states.   
In: VIIIth International Congress of the International Association of Mathematical Physics, 
Mebkhout M., Seneor, R. (Eds.), 
World Scientific, Singapore, 1987, pp. 331-337. 
Arxiv, 
http://arxiv.org/abs/1206.3734  

Lebowitz, J.L., Pfister, C.E. (1981):
Surface tension and phase coexistence,
Phys. Rev. Lett. {\bf 46}, 1031-1033. 

Messager, A., Miracle-Sole, S., Ruiz, S. (1992): 
Convexity properties of the surface tension and equilibrium crystals, 
J. Stat. Phys. {\bf 67}, 449-470.  

Miracle-Sole, S., Ruiz, J. (1994): 
On the Wulff construction as a problem of equivalence of ensembles.  
In: On Three Levels: Micro, Meso and Macroscopic Approaches in Physics, 
Fannes, M., Verbeure, A. (Eds.), pp. 295-302,      
Plenum Press, New York.  
Arxiv, 
http://arxiv.org/abs/1206.3739 

Miracle-Sole, S. (1995): 
Surface tension, step free energy and facets in the
equilibrium crystal shape, 
J. Stat. Phys. {\bf 79}, 183-214. 

Miracle-Sole, S. (1999):
Facet shapes in a Wulff crystal.  
In: Mathematical Results in Statistical Mechanics, 
Miracle-Sole S., Ruiz J., Zagrebnov, V., eds., 
pp. 83-101,  
World Scientific, Singapore. 
Arxiv, 
http://arxiv.org/abs/1206.3736 

Nienhuis, B., Hilhorst, H.J., Bl\"ote, H.B. (1984):
Triangular SOS models and cubic crystal shapes, 
J. Phys. A: Mat. Gen. {\bf 17}, 3559-3581. 

Pfister, C.E. (1991):
Large deviations and phase separation in the two-di\-men\-sio\-nal Ising model, 
Helv. Phys. Acta {\bf 64}, 953-1054. 

Schneer, C.J. (1977):
Crystal Form and Structure,  
Benchmark papers in Geology, vol. 34, 
Dowden, Hutchinson and Ross, Stroudsbourg (USA).

Taylor, J.E. (1987): 
Some crystalline variational techniques and results,  
Asterisque, 
vol. 154, pp. 307-320.  

Wolf, P.E., Balibar, S., Gallet, F. (1983): 
Experimental observations of a third roughening 
transition in hcp $^4$He crystals, 
Phys. Rev. Lett. {\bf 51}, 1366-1369. 

\end{document}